\documentclass{iau}
\usepackage{graphicx}

\title[The evolution of inner disk radius in Circinus X-1] 
{The evolution of inner disk radius with orbital phase in Circinus X-1}

\author[Yanan Wang, Guoqiang Ding \& Chunping Huang]   
{Yanan Wang$^{1,2}$, Guoqiang Ding$^1$, 
\and Chunping Huang$^{1,2}$}

\affiliation{$^1$Xinjiang Astronomical Observatory, Chinese Academy of Sciences, 
\\  150, Science 1-Street, Urumqi, Xinjiang 830011, China 
\\ email: {\tt wangyanan@xao.ac.cn, dinggq@gmail.com} \\[\affilskip]
$^2$University of Chinese Academy of Sciences, \\ China}

\pubyear{2012}
\volume{290}  
\jname{Feeding compact objects: Accretion on all scales}
\editors{C.M. Zhang, T. Belloni, M. M\'endez \& S.N. Zhang, eds.}

\begin{document}

\maketitle

\begin{abstract}

Using {\it RXTE} observations, we investigate the evolution of inner disk 
radius ($R_{\rm in}$) of Cir X-1 during two cycles and find obvious orbital 
modulation. We argue that the modulation is attributed to its high orbital 
eccentricity. The disk luminosity is inversely with the inner disk temperature 
($kT_{\rm in}$), which is ascribed to the slow increase of $kT_{\rm in}$ and, however, 
the rapid decrease of $R_{\rm in}$ during the passage for the neutron star to 
depart from the companion star.

\end{abstract}

\firstsection 
\section{Introduction}

Circinus X-1 (Cir X-1), with an orbital period of ~16.6 days (\cite[Kaluzienski 
et al. 1976]{Kaluzienski1976}) and a distance of 5.5 kpc (\cite[Case \& 
Bhattacharya 1998]{Case1998}), is a X-ray binary. Its compact star has been 
considered as a neutron star (NS) since the discovery of type-I X-ray bursts in 
this source (\cite[Tennant et al. 1986]{Tennant1986}). One of the special 
parameters of Cir X-1 is its high orbital eccentricity ($\sim$ 0.7-0.9) 
(\cite[Johnston et al. 1999]{Johnston1999}), which makes it a peculiar source. 
\cite[Shirey et al. (1996)]{Shirey1996} found the orbital modulation for the
spectrum and quasi-periodic oscillation (QPO) of Cir X-1 
and \cite[Ding et al. (2006b)]{Ding2006b} found that its power-law (PL) 
hard tail, evolving on its hardness-intensity diagram (HID) (\cite[Ding et al. 
2003]{Ding2003}), is modulated by orbital phase too. At the periastron, a long-term 
dip was present (\cite[Ding et al. 2006a]{Ding2006a}). For explaining the 
behaviors of observed optical and infrared (IR) emission lines, 
\cite[Johnston et al. (1999)]{Johnston1999} proposed that the accretion and 
accretion disk of this NS X-ray binary (NSXB) could evolve with orbital phase. 

\section{Data Analysis}

With software HEASOFT 6.11 and FTOOLS V.6.11, we choose the observations during 
two orbital periods (1996 September 21--October 7, 1996 March 8--19) of Cir X-1 
to perform our analysis. Following \cite [Stewart et al. (1991)]{Stewart1991}, 
the time of zero phase is given by the ephemeris equation
\begin{equation}
JD_0 = 2443076.87+(16.5768-0.0000353N)N.
\end{equation}
We produce the background-subtracted PCA spectra at different phases. 
\cite[Shirey et al. (1999)]{Shirey1999} used several spectral models to fit the 
PCA spectra of Cir X-1 and found that the best-fit model is the so-called 
Eastern model, consisting of a blackbody (BB) and a multicolor disk blackbody 
(MCD), which are interpreted as the emission from the NS surface and the 
optically thick accretion disk, respectively. We adopt this model, use it to fit 
the spectra of Cir X-1 during the two orbital periods, and then get the inner 
disk temperature $(T_{\rm in})$, inner disk radius ($R_{\rm in}$), BB temperature 
$(T_{\rm bb})$, and BB radius $(R_{\rm bb})$. The BB radii multiplied by a 
coefficient are considered as the typical NS radius.   

\section{Result and Discussion}

\begin{figure}[b]
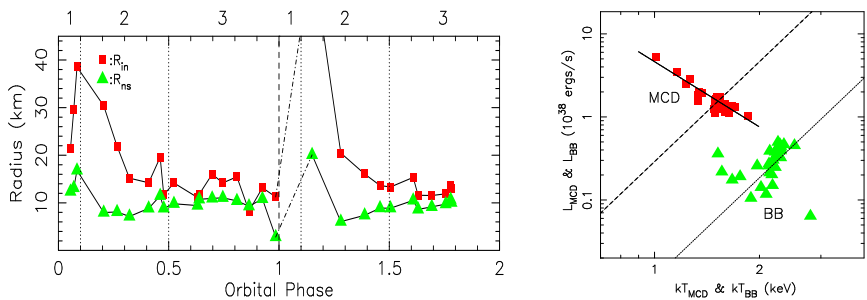

\begin{center}
 \includegraphics[width=1.5in,height=2.6in,angle=-90]{wangy_fig1_1.ps}
 \hspace{0.5cm}  
 \includegraphics[width=1.5in,height=1.6in,angle=-90]{wangy_fig1_2.ps}
 \caption{Left panel: the evolution of inner disk radius and the inferred NS 
radius along orbital phase. Right panel: the luminosities of spectral 
components (MCD/BB) vs. their characteristic temperatures; the dashed line and 
dotted line correspond to $L=4\pi R^2\sigma T^4$, with $R=15\ {\rm km}$ 
and $R=3\ {\rm km}$, respectively; the solid line corresponds 
to $L\propto T^{-2.6}$.}
   \label{fig1}
\end{center}
\end{figure}

As shown in the right panel of Figure~\ref{fig1}, the disk emission deviates from 
the relation of $L\propto T^4$, which indicates that $R_{\rm in}$ is varied, because 
of $L_{\rm disk}=4\pi R^2_{\rm in}\sigma T^4_{\rm in}$. In panel A of 
Figure~\ref{fig1}, one can see that at the periastron (phase 0-0.1) $R_{\rm in}$ 
increases abruptly, then from phase 0.1 to the apastron (phase 0.5) $R_{\rm in}$ 
decreases rapidly, and, finally, from the apastron to phase 1 the $R_{\rm in}$ roughly 
steadies. As suggested by \cite[Johnston et al. (1999)]{Johnston1999}, at the 
periastron the large tidal force could make the disk unstable, resulting in large 
variation of $R_{\rm in}$; after the apastron until the apastron the disk is 
formed gradually and meanwhile the NS departs from the companion star, but the 
disk moves towards the NS due to decrease of radiation pressure; after the 
apastron, steady accretion takes place on the formed disk. It is obvious that during 
the passage for the NS to leave the companion star the slow increase 
of $(T_{\rm in})$ and, however, the rapid decrease of $R_{\rm in}$ contribute to the 
inverse correlation between the disk luminosity and the inner disk temperature. 

\section{Acknowledgements}

This work is supported by the National Basic Research Program of China (973 Program 
2009CB824800) and the Natural Science Foundation of China under grant no. 11143013.

\end{document}